\documentclass[11pt]{article}

\newcommand{\jpa}{J. Phys. A~}

\newcommand{\natphy}{Nature Phys.~}

\newcommand{\prl}{Phys. Rev. Lett.~}
\newcommand{\pra}{Phys. Rev. A~}
\newcommand{\pla}{Phys. Lett. A~}

\usepackage{algorithmic}
\usepackage{young}
\usepackage[latin1]{inputenc}
\usepackage[linesnumbered,boxed,ruled,commentsnumbered]{algorithm2e}
\usepackage{appendix}
\usepackage{epstopdf}
\usepackage{xcolor}
\usepackage{subfigure}
\usepackage[T1]{fontenc}
\usepackage[sc]{mathpazo}
\usepackage{amsmath}
\usepackage{amssymb}
\usepackage{graphicx,color}
\usepackage{framed}
\usepackage{multirow}
\usepackage{enumerate}
\usepackage{amsthm}
\usepackage{amsfonts,mathrsfs}
\usepackage{geometry} 
\usepackage{eepic}
\usepackage{ifthen}
\usepackage[vcentermath]{youngtab}
\usepackage[unicode=true,pdfusetitle, bookmarks=true,bookmarksnumbered=false,bookmarksopen=false, breaklinks=false,pdfborder={0 0 0},backref=false,colorlinks=false] {hyperref}
\hypersetup{
colorlinks,linkcolor=myurlcolor,citecolor=myurlcolor,urlcolor=myurlcolor}
\definecolor{myurlcolor}{rgb}{0,0,0.7}

\usepackage{color}

\newcommand{\blue}{\textcolor{blue}}

\newcommand{\proj}[1]{| #1\rangle\!\langle #1 |}

\usepackage{hyperref}
\hypersetup{pdfpagemode=UseNone}

\newcommand{\tinyspace}{\mspace{1mu}}

\newcommand{\abs}[1]{\left\lvert\tinyspace #1 \tinyspace\right\rvert}

\newcommand{\norm}[1]{\left\lVert\tinyspace #1 \tinyspace\right\rVert}

\newcommand{\setft}[1]{\mathrm{#1}}

\newcommand{\density}[1]{\setft{D}\left(#1\right)}

\def\vol{\mathrm{vol}}

\def \dif {\mathrm{d}}

\def \vol {\mathrm{vol}}

\def\complex{\mathbb{C}}

\def\I{\mathbb{1}}

\newenvironment{mylist}[1]{\begin{list}{}{
    \setlength{\leftmargin}{#1}
    \setlength{\rightmargin}{0mm}
    \setlength{\labelsep}{2mm}
    \setlength{\labelwidth}{8mm}
    \setlength{\itemsep}{0mm}}}
    {\end{list}}

\def\ot{\otimes}

\newcommand{\out}[2]{| #1\rangle\langle #2 |}

\newcommand{\Pa}[1]{\left(#1\right)}

\newcommand{\Br}[1]{\left[#1\right]}

\newcommand{\Set}[1]{\left\{#1\right\}}

\newcommand{\bra}[1]{\langle#1|}

\newcommand{\ket}[1]{|#1\rangle}

\DeclareMathOperator{\trace}{Tr}

\newcommand{\Ptr}[2]{\trace_{#1}\Pa{#2}}

\newcommand{\Tr}[1]{\Ptr{}{#1}}

\def\cH{\mathcal{H}}
\def\cN{\mathcal{N}}

\def\bsU{\boldsymbol{U}}

\def\bsp{\boldsymbol{p}}

\def\rB{\mathrm{B}}
\def\rG{\mathrm{G}}\def\rH{\mathrm{H}}

\def\rW{\mathrm{W}}\def\rX{\mathrm{X}}
\def\rZ{\mathrm{Z}}

\def\B{\textsf{B}}

\def\bbC{\mathbb{C}}

\newtheorem{thrm}{Theorem}[section]

\newtheorem{prop}[thrm]{Proposition}
\newtheorem{cor}[thrm]{Corollary}
\theoremstyle{definition}

\newtheorem{exam}[thrm]{Example}

\numberwithin{equation}{section}

\newcounter{questionnumber}

\begin{document}

\title{\bf\large A characterization of entangled two-qubit states via partial-transpose-moments}

\author{\blue{Lin Zhang}$^1$\footnote{E-mail: godyalin@163.com},\,\, \blue{Ming-Jing Zhao}$^2$,\,\, \blue{Lin Chen}$^3$\footnote{E-mail: linchen@buaa.edu.cn},\,\, \blue{Hua Xiang}$^4$,\,\, \blue{Yi Shen}$^5$\footnote{E-mail: yishen@jiangnan.edu.cn}\\
  {\it\small $^1$ School of Science, Hangzhou Dianzi University, Hangzhou 310018, China}\\
  {\it\small $^2$ School of Science, Beijing Information Science and Technology University, Beijing, 100192, China}\\
  {\it\small $^3$ School of Mathematical Sciences, Beihang University, Beijing 100191, China}\\
  {\it\small $^4$ School of Mathematics and Statistics, Wuhan University, Wuhan 430072, China}\\
  {\it\small $^5$ School of Science, Jiangnan University, Wuxi, Jiangsu 214122, China}}
\date{}
\maketitle

\begin{abstract}
Although quantum entanglement is an important resource, its
characterization is quite challenging. The partial transposition is
a common method to detect bipartite entanglement. In this paper, the
authors study the partial-transpose(PT)-moments of two-qubit states,
and completely describe the whole region, composed of the second and
third PT-moments, for all two-qubit states. Furthermore, they
determine the accurate region corresponding to all entangled
two-qubit states. The states corresponding to those boundary points
of the whole region, and to the border lines between separable and
entangled states are analyzed. As an application, they characterize
the entangled region of PT-moments for the two families of Werner
states and Bell-diagonal states. The relations between entanglement
and the pairs of PT-moments are revealed from these typical
examples. They also numerically plot the whole region of possible
PT-moments for all two-qubit X-states, and find that this region is
almost the same as the whole region of PT-moments for all two-qubit
states. Moreover, they extend their results to detect the
entanglement of multiqubit states. By utilizing the PT-moment-based
method to characterize the entanglement of the multiqubit states
mixed by the GHZ and W states, they propose an operational way of
verifying the genuine entanglement in such states.
\end{abstract}
\newpage
\tableofcontents
\newpage

\section{Introduction}

Characterizing entanglement is a key task in quantum information
science because the entanglement is a necessary resource in various
information processing tasks
\cite{1Bennet1992,2Bennet1993,3Pan1997,4Deutsch1996}. To detect
entanglement, Peres firstly proposed a necessary condition for all
separable states \cite{5Peres1996}, i.e., the well-known
positive-partial-transpose (PPT) criterion. It states that a
bipartite state is entangled whenever it violates the PPT criterion,
i.e., its partial transpose (PT) has at least one negative
eigenvalue. Such entangled states are called
non-positive-partial-transpose (NPT) states. Subsequently,
Horodecki's family specified that the PPT criterion is a necessary
and sufficient condition \cite{6Horodecki1996pla} for the separable
states supported on $\complex^m\ot\complex^n$ when $mn\leqslant 6$.
Due to the PPT criterion, the entanglement detection has been
reduced to detect the entangled states retaining the PPT property,
also known as the PPT entangled states. A fundamental tool to detect
PPT entangled states is the so-called entanglement witness (EW)
\cite{7Terhal2000}. It has been shown that an EW can detect PPT
entangled states if and only if it is non-decomposable
\cite{8Lewenstein2000}. Thus much effort has been devoted to
constructing non-decomposable EWs
\cite{9Skowronek2009,10Chruscinski2010,11Ha2011}. In particular, the
partial transpose of an NPT state can be viewed as an EW, though it
can only detect NPT states.

The PPT criterion implies that the negative eigenvalues of the
partial transpose of a bipartite state $\rho_{AB}$ are a signature
of the entanglement in $\rho_{AB}$. Denote by $\rho_{AB}^{\Gamma_A}$
and $\rho_{AB}^{\Gamma_B}$ the partial transposes of $\rho_{AB}$
with respect to system $A$ and $B$ respectively. Note that the two
partial transposes share the same spectrum. Hence, for simplicity,
we use $\rho_{AB}^{\Gamma}$ to denote one of the two partial
transposes. Depending on the spectrum of $\rho^{\Gamma}_{AB}$, a
well-known computable entanglement measure called the logarithm of
negativity was introduced in Ref. \cite{12Plenio2005}. It has
various operational interpretations, including the upper bound of
the entanglement distillation rate \cite{13Horodecki2000}, the bound
on teleportation capacity and the entanglement cost under a large
operation set \cite{14Vidal2002,15Audenaert2003}.

The PPT criterion is actually a mathematical tool commonly used in
the theoretical scenario. For a given bipartite state $\rho_{AB}$ of
system $AB$, it is straightforward to check whether the PPT
criterion is violated. Nevertheless, in the practical scenario, the
detection of entanglement is not so direct. On the one hand, the
partial transposition of an operator is not a physical operation,
and it cannot be implemented directly in experiments. On the other
hand, in actual experiments, the quantum state is unknown, unless
the resource-inefficient quantum state tomography is performed
\cite{16Yu2021}. More generally, the realizations of effective
entanglement criteria usually consume exponential resources, and
efficient criteria often perform poorly without prior knowledge
\cite{17Liu2022}. Therefore, it is necessary to develop
experimentally efficient methods to detect entanglement based on the
measurable observables. Recently, a realizable entanglement
detection method based on the PPT criterion by investigating the
so-called \emph{partial transpose moments} (PT-moments) defined by
\begin{eqnarray}\label{eq:defptm}
p_k := \Tr{\Br{\rho^{\Gamma}_{AB}}^k}
\end{eqnarray}
has been introduced in Ref. \cite{16Yu2021}. Although the PT-moments
are nonlinear functions of the state $\rho_{AB}$ depending on the
spectrum of $\rho^{\Gamma}_{AB}$, it has been shown, recently, in
Refs. \cite{18Elben2020,19Huang2020} that the PT-moments can be
efficiently measured from local randomized measurements.
Specifically, the PT-moment can be measured as an expectation value
of an $n$-copy cyclic permutation operator \cite{18Elben2020}, and
the protocol of measurements only requires single-qubit control in
Noisy Intermediate Scale Quantum (NISQ) devices \cite[Fig.
1b]{18Elben2020}. Furthermore, Ref. \cite{20Gray2018} shows that
with the assistance of machine learning, one can extract the
negativity just from the third PT-moment $p_3$. The latter can be
estimated based on the random unitary evolution and local
measurements on a single-copy quantum state \cite{21Zhou2020}.
Later, Liu \emph{et al.} extended the notion of PT-moments to that
of permutation moments \cite{22Liu2022}, and proposed a framework
for designing multipartite entanglement criteria based on
permutation moments. Moreover, the PT-moments provide an
entanglement detection method via the moments of the state
$\rho_{AB}$ itself. Not only this, there are several proposals of
testing and characterizing entanglement via second-order and special
higher-order moments of various observables (like the position,
momentum, annihilation and creation operators of qubits and qudits
\cite{23Simon2000,24Duan2000,25Mancini2002,26Agarwal2005}) has been
presented, and their matrices have been developed by Werner Vogel et
al. for both bipartite \cite{27Shchukin2005,28Miranowicz2006} and
multipartite (or multimode) \cite{29Vogel2008,30Miranowicz2010}
systems.

The measurability of PT-moments bridges the practical limitations of
the PPT criterion, and allows it to be used for experimental
entanglement detection. In spite of this, the PT-moments cannot
provide a more powerful criterion than the PPT criterion for
entanglement detection. On the other word, we cannot detect the PPT
entangled states based on the knowledge of PT-moments. Moreover, we
cannot use PT-moments to distinguish the set of PT-moments of all
separable states $\sigma_{AB}$ from that of all states $\rho_{AB}$
on the same underlying space because both of the two sets are the
same. Hence, in order to dectect entanglement utilizing the
PT-moments, it is necessary to characterize the PT-moments of all
NPT entangled states. Basically, we consider the characterization of
the set of PT-moments of all entangled states supported on
$\complex^m\ot\complex^n$, where $m,n\geqslant2$. In such a
framework, the computational complexity increases exponentially as
$m$ and $n$ increase. Due to the fundamental importance of the toy
model of the two-qubit system, we focus on the derivation in the
two-qubit system. Since there is no PPT entangled state of the
two-qubit system, the PT-moment-based method has the ability to
characterize the set of all two-qubit entangled states.
Proposition~\ref{prop:yu} is constructed and can be viewed as the
identification of the set of PT-moments of all two-qubit separable
states. In this short report, we will identify the set of PT-moments
of all entangled two-qubit states in Theorem \ref{th:A1}. By doing
so, in fact, we give an equivalent, but experimentally measurable
criterion of entangled two-qubit states. Such a criterion is also
illustrated in Fig. \ref{ptregion} by identifying the dividing line
between the set of separable states and that of entangled states.
Then we apply this experimentally measurable criterion to detect the
entanglement of several widely used states. We also demonstrate the
relations between entanglement and the pairs of PT moments for these
typical examples in Figs. \ref{bdregion} - \ref{nqubit}. Analogously
to these figures, several proposals of characterizing the
entanglement of two-qubit states by plotting one entanglement
measure or witness versus others have been reported in a number of
interesting papers
\cite{31Wei2003,32Bartkiewicz2013,33Horst2013,34Bartkiewicz2015}.

From Eq. \eqref{eq:defptm} one can equivalently calculate each $p_k$
as the sum of all $x_i^k$, where $x_i$'s are the eigenvalues of
$\rho_{AB}^{\Gamma}$. Thus, the calculation of $p_k$ is closely
related to the spectrum of $\rho^{\Gamma}_{AB}$. Due to this close
relation, it is necessary to deeply understand some properties of
the spectrum of $\rho^{\Gamma}_{AB}$. Recently, several results on
the spectrum of $\rho^{\Gamma}_{AB}$ have been developed
\cite{35Rana2013,36Johnston2013,37Shen2020,38Duan2022}.
Specifically, it follows from Ref. \cite{35Rana2013} that the
partial transpose $\rho^{\Gamma}_{AB}$ supported on
$\complex^m\ot\complex^n$ has no more than $(m-1)(n-1)$ negative
eigenvalues, and all eigenvalues of $\rho^{\Gamma}_{AB}$ fall within
the closed interval $\Br{-\tfrac12,1}$. Therefore, when considering
the PT-Moment problem for bipartite space $\complex^2\ot\complex^3$,
we will encounter more difficulties of calculating the PT-moments
because the number of negative eigenvalues of $\rho^{\Gamma}_{AB}$
could be one or two. In view of this, the identification of all
PT-moments of the states supported on $\complex^m\ot\complex^n$ with
$mn\geqslant6$ could be much more complicated than the case of
two-qubit states.

The paper is organized as follows. In Sec. \ref{sect:mainresult}, we
present our main result with its proof being delayed to Appendix
\ref{app:A1}. In Sec. \ref{examples}, some examples of two-qubit and
multiqubit states are given in terms of our main result. Finally, we
conclude in Sec. \ref{sec:con}.

\section{Preliminaries}

To see how the PT-moments can be used in entanglement detection,
suppose that $\rho_{AB}$ is a bipartite state supported on the
Hilbert space $\cH_A\ot\cH_B\cong \bbC^{d_A}\ot\bbC^{d_B}$, and we
know the PT-moment $p_k$ of $\rho_{AB}$ for any $k\in[1,d]$, where
$d=d_Ad_B$ is the global dimension. Denote by $(x_1,\ldots,x_d)$ the
spectrum of $\rho^{\Gamma}_{AB}$, where the eigenvalues
$x_1,\ldots,x_d$ are always assumed to be sorted in descending
order, unless otherwise stated. Obviously, the PT-moment for each
$k$ is formulated as
\begin{eqnarray}\label{eq:pt-spectrum}
p_k=\sum_{i=1}^d x_i^k.
\end{eqnarray}
Then Ref. \cite{39Ekert2002} shows that the spectrum of
$\rho^{\Gamma}_{AB}$ can be determined with all known PT-moments,
for which whether $\rho_{AB}$ is an NPT state can be verified. This
process provides the PT-moment-based entanglement detection.
Hereafter, we call $\bsp^{(d)}=(p_1,\ldots,p_d)$ the PT-moment
vector, and always assume that $p_1=1$ for convenience.

In Ref. \cite{16Yu2021}, the authors studied a specific problem as
follows.

\noindent{\bf PT-Moment problem:} Given the PT-moments of order $n$,
is there a separable state compatible with the data? More
technically formulated, \emph{given} the real vector
\begin{eqnarray}
\bsp^{(n)}=(p_1,\ldots,p_n),
\end{eqnarray}
\emph{is there} a separable quantum state $\rho_{AB}$ such that for
each $1\leq k\leq n$,
\begin{eqnarray}
p_k = \Tr{\Br{\rho^{\Gamma}_{AB}}^k}?
\end{eqnarray}
It is natural to consider the detection of entanglement in
$\rho_{AB}$ from a few of the PT-moments due to the difficulty in
measuring all the PT-moments. In fact, the authors derived a
necessary and sufficient condition for the PT-Moment problem when
the order $n$ is three in Ref. \cite{16Yu2021}. It states that there
is a separable state supported on $\bbC^d$ compatible with the given
vector $(q_1,q_2,q_3)$ if and only if
\begin{eqnarray}\label{eq:knownineq-1}
\begin{aligned}
&\frac{1}{d}\leqslant q_2 \leqslant 1,\\
&\Phi^+_d(q_2)\geqslant q_3\geqslant \phi_d(q_2),
\end{aligned}
\end{eqnarray}
where
\begin{eqnarray}
\label{eq:knownineq-2}
\begin{aligned}
\Phi^+_d(q_2)&:=(d-1)y^3+[1-(d-1)y]^3,\\
\phi_d(q_2)&:=\tau(q_2)x^3+(1-\tau(q_2)x)^3,
\end{aligned}
\end{eqnarray}
for $\tau(q_2):=\lfloor\frac1{q_2}\rfloor$, and via
\begin{eqnarray}
\label{eq:knownineq-3}
\begin{aligned}
y&\equiv y(q_2):=\frac{d-1-\sqrt{(d-1)(dq_2-1)}}{d(d-1)},\\
x&\equiv
x(q_2):=\frac{\tau(q_2)+\sqrt{\tau(q_2)[(\tau(q_2)+1)q_2-1]}}{\tau(q_2)(\tau(q_2)+1)}.
\end{aligned}
\end{eqnarray}
Based on this result and in view of the result from Ref.
\cite{40Horodecki1996}, we get a more precise answer to the
PT-Moment problem for two-qubit states.
\begin{prop}\label{prop:yu}
There is a separable two-qubit state $\rho_{AB}$ compatible with the
given PT-moment vector $\bsp^{(4)}=(p_1,p_2,p_3,p_4)$, where every
PT moment $p_k$ is defined by Eq. \eqref{eq:defptm}, if and only if
the following two inequalities hold:
\begin{eqnarray}
\label{eq:propsep-1}
&&\frac14 \leqslant p_2 \leqslant 1,\\
\label{eq:propsep-2} &&\Phi^+_4(p_2) \geqslant p_3 \geqslant
\phi_4(p_2),
\end{eqnarray}
where $\Phi^+_4(p_2)=\frac{3(6p_2-1)+
\sqrt{3}(4p_2-1)^{\frac32}}{24}$, and $\phi_4(p_2)$ can be rewritten
as a piecewise function:
\begin{eqnarray}
\label{eq:prop1-1} \phi_4(p_2) =
\begin{cases}
\frac{3 p_2-1}2,&\text{if }p_2\in[\frac12,1],\\
\frac{2(9p_2-2)-\sqrt{2}(3p_2-1)^{\frac32}}{18},&\text{if
}p_2\in[\frac13,\frac12],\\
\frac{3(6p_2-1)- \sqrt{3}(4p_2-1)^{\frac32}}{24},&\text{if
}p_2\in[\frac14,\frac13].
\end{cases}
\end{eqnarray}
This is a continuous function for $p_2\in[\frac14,1]$.
\end{prop}

\section{Main result}\label{sect:mainresult}

In this section we propose a PT-moment-based separability criterion
for two-qubit states. It relies on the accurate characterization of
the set of the PT-moment vectors corresponding to entangled
two-qubit states as follows.
\begin{thrm}\label{th:A1}
There is an entangled two-qubit state $\rho_{AB}$ compatible with
the given PT-moment vector $\bsp^{(4)}=(p_1,p_2,p_3,p_4)$, where
every PT moment $p_k$ is defined by Eq. \eqref{eq:defptm}, if and
only if both the following two inequalities hold:
\begin{eqnarray}
\label{eq:mainthm-1}
&&\frac13 < p_2\leqslant 1,\\
\label{eq:mainthm-2} &&\phi_4(p_2)>p_3\geqslant \Phi^-_4(p_2),
\end{eqnarray}
where $\phi_4(p_2)$ is given by Eq. \eqref{eq:prop1-1}, and
\begin{equation}
\label{eq:Phi4-}
\Phi^-_4(p_2):=\frac{3(6p_2-1)-\sqrt{3}(4p_2-1)^{\frac32}}{24}.
\end{equation}
\end{thrm}
Note here that $\phi_4(p_2)\equiv\Phi^-_4(p_2)$ for all
$p_2\in[\frac14,\frac13]$. This amounts to say that the two-qubit
state $\rho_{AB}$ with purity $p_2\in[\frac14,\frac13]$ must be
separable.
\begin{proof}
The details are put in Appendix~\ref{app:A1}.
\end{proof}

\begin{figure}[ht]\centering
{\begin{minipage}[b]{1\linewidth}
\includegraphics[width=1\textwidth]{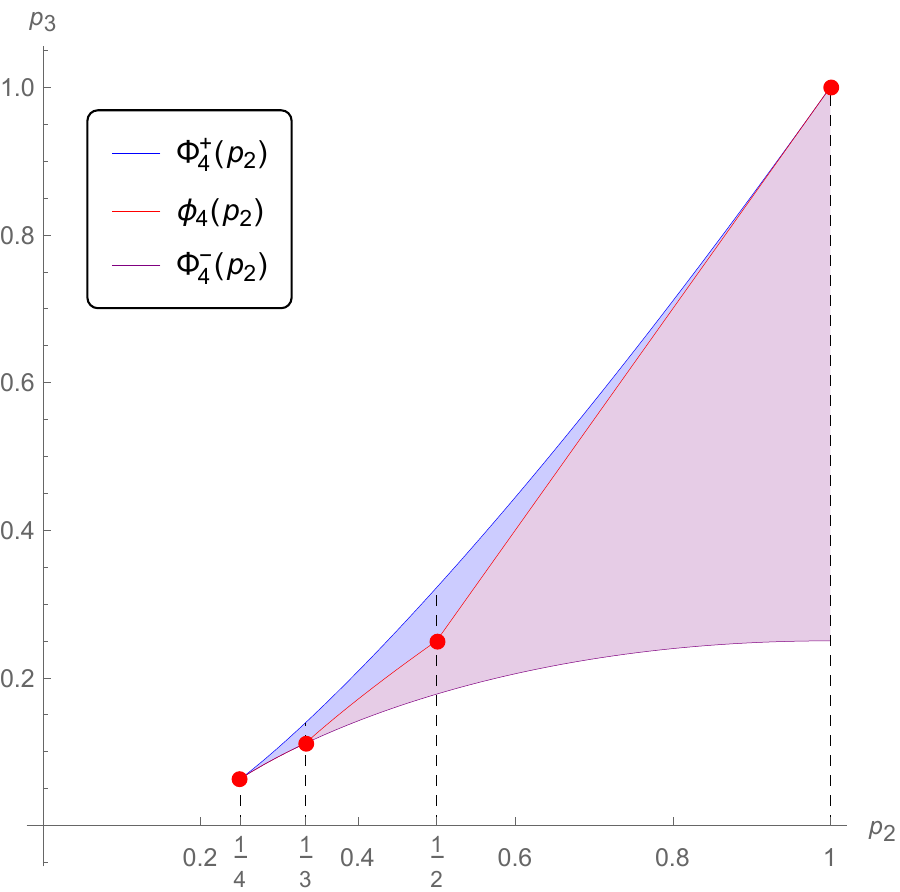}
\end{minipage}}
\caption{(Color Online) The PT-moment pair $(p_2,p_3)$ of two-qubit
state $\rho_{AB}$ falls within the light blue colored region with
upper and lower boundaries iff two-qubit $\rho_{AB}$ is separable;
the pair $(p_2,p_3)$ of PT-moments of two-qubit state $\rho_{AB}$
falls within the purple colored region with only the lower boundary
iff two-qubit $\rho_{AB}$ is entangled. The coordinates of these red
points (from left to right) correspond to $(\frac14,\frac1{16})$(
only the completely mixed state), $(\frac13,\frac19),
(\frac12,\frac14),(1,1)$(the set of separable pure states),
respectively; the coordinate of only the purple point is
$(1,\frac14)$, corresponding to the family of all maximally
entangled states.}\label{ptregion}
\end{figure}
Based on the characterization of the set of PT-moment vectors of
entangled two-qubit states, we further obtain that of all two-qubit
states.
\begin{cor}
For any two-qubit state $\rho_{AB}$, its PT-moment vector
$\bsp^{(4)}=(p_1,p_2,p_3,p_4)$ satisfies that
\begin{eqnarray}
\Phi^+_4(p_2)\geqslant p_3\geqslant \Phi^-_4(p_2),
\end{eqnarray}
where $\Phi^\pm_4(p_2)=\frac{3(6p_2-1)\pm
\sqrt{3}(4p_2-1)^{\frac32}}{24}$ for $p_2\in[\frac14,1]$.
\end{cor}

Denote by $R_{\text{sep}}$ the set of pairs $(p_2,p_3)$ of
PT-moments of all separable two-qubit states; and by
$R_{\text{ent}}$ that of all entangled two-qubit states. That is,
\begin{eqnarray}\label{eq:defentsepset}
\begin{aligned}
R_{\text{sep}}&:=\Set{(p_2,p_3): \tfrac14\leqslant
p_2\leqslant1,\Phi^+_4(p_2)\geqslant p_3\geqslant \phi_4(p_2)},\\
R_{\text{ent}}&:=\Set{(p_2,p_3): \tfrac14\leqslant
p_2\leqslant1,\phi_4(p_2)> p_3\geqslant \Phi^-_4(p_2)}.
\end{aligned}
\end{eqnarray}
Both of them are plotted in Fig~\ref{ptregion}. In addition, we
present specific states corresponding to boundary curves of
$R_{\mathrm{sep}/\mathrm{ent}}$ in Fig~\ref{ptregion}:
\begin{itemize}
\item The vertical line $p_2=1$: All pure states.
\item The curve $p_3=\Phi^+_4(p_2)$:
\begin{eqnarray}
\lambda(\out{00}{00}+\out{01}{01}+\out{10}{10})+(1-3\lambda)\out{11}{11}
\end{eqnarray}
for $\lambda\in[0,\frac14]$.
\item The curve $p_3=\Phi^-_4(p_2)$: All Werner states.
\end{itemize}
Specific states corresponding to the inner border lines:
\begin{itemize}
\item The curve $p_3=\phi_4(p_2)(p_2\in[\tfrac12,1])$:
\begin{eqnarray}
\lambda\out{00}{00}+(1-\lambda)\out{01}{01}
\end{eqnarray}
for $\lambda\in[0,1]$.
\item The curve $p_3=\phi_4(p_2)$ with $p_2\in[\tfrac13,\tfrac12]$:
\begin{eqnarray}
\lambda\out{00}{00}+\lambda\out{01}{01}+(1-2\lambda)\out{10}{10}
\end{eqnarray}
for $\lambda\in[\tfrac13,\tfrac12]$.
\item The curve $p_3=\phi_4(p_2)$ with $p_2\in[\tfrac14,\tfrac13]$: All separable Werner states.
\end{itemize}

Furthermore, we also calculate the areas of two regions:
$A(R_{\text{sep}}) = \frac{43}{2160}$ and $A(R_{\text{ent}}) =
\frac{443}{2160}$. There is an interesting question may be related
to such two areas. That is, we may establish the joint probability
density function (PDF)
\begin{eqnarray}
f(p_2,p_3):=\int\dif\mu(\rho)\delta\Pa{p_2 -
\Tr{\Br{\rho^\Gamma}^2}}\delta\Pa{p_3 - \Tr{\Br{\rho^\Gamma}^3}}
\end{eqnarray}
over the set of pairs $(p_2,p_3)$ for all two-qubit states, where
$\dif\mu(\rho)$ is the measure induced by Hilbert-Schmidt norm. Once
we get the analytical expression of function $f$, then we may solve
the following open question. That is,
\begin{eqnarray}
\int_{R_{\mathrm{sep}}}f(p_2,p_3)\dif p_2\dif
p_3\Big/\int_{R_{\mathrm{ent}}}f(p_2,p_3)\dif p_2\dif p_3
\end{eqnarray}
should be related to the ratio
$\frac{\vol_{\mathrm{sep}}(\density{\complex^2\ot\complex^2})}{\vol_{\mathrm{ent}}(\density{\complex^2\ot\complex^2})}$,
which is conjectured to be $\tfrac8{25}$ \cite{41Lovas2017}.
Hopefully, it sheds new light on this open question.

\section{Application}\label{examples}

In this section, as essential applications of our main result in
Sec. \ref{sect:mainresult}, we  construct examples, and use the
PT-moment-based separability criterion in Theorem \ref{th:A1} to
analyze the separability of these examples. We construct examples of
the two-qubit system in Sec. \ref{subsec:twoqubit}, such as the
Werner states, Bell-diagonal states, $X$-states. Then we also
consider the multipartite entanglement, and construct examples of
the multipartite system in Sec. \ref{subsec:multi}, such as the
convex sum of multiqubit GHZ and W states.

\subsection{two-qubit systems}\label{subsec:twoqubit}

\begin{exam}[The family of Werner states]
In 1989, Werner analytically constructed a family of $\bsU\ot\bsU$
invariant states to investigate local hidden variable (LHV) models
\cite{42Werner1989}. As a toy model, we consider the two-qubit
Werner states formulated as
\begin{eqnarray}\label{eq:defwerners}
\rho_W(w) = w\out{\psi^-}{\psi^-}+(1-w)\frac{\mathbf{1}_4}4
\end{eqnarray}
where $\ket{\psi^-}=\frac{\ket{01}-\ket{10}}{\sqrt{2}}$ and
$w\in[0,1]$. By calculation, the second and third PT-moments of
two-qubit Werner state $\rho_W(w)$ are, respectively,
\begin{eqnarray}\label{eq:wernerptm-1}
p_2=\frac{1+3w^2}4\quad\text{and}\quad p_3=\frac{-6 w^3+9
w^2+1}{16}.
\end{eqnarray}
Substituting $p_2$ and $p_3$ above into the two inequalities
\eqref{eq:mainthm-1} and \eqref{eq:mainthm-2} in
Theorem~\ref{th:A1}, one can verify both the two inequalities hold
if and only if $w\in(\tfrac13,1]$. This is also equivalent to the
well-known fact on the separability of Werner states
\cite{43Divincenzo2000} that $\rho_W(w)$ is entangled if and only if
$w\in(\frac{1}{3},1]$. Eliminating the parameter $w\in[0,1]$, we
find that the pair $(p_2,p_3)$ satisfies $p_3=\Phi^-_4(p_2)$. It
means the points fixed by the pairs $(p_2,p_3)$ corresponding to the
set of entangled two-qubit Werner states are just lying on the
lowest curve, i.e., the purple one in Fig~\ref{ptregion}.
\end{exam}

\begin{exam}[The family of Bell-diagonal states]
The Bell-diagonal states \cite{40Horodecki1996} in the two-qubit
system can be written as
\begin{eqnarray}
\label{eq:defbelldiag} \rho_{\text{Bell}}=\frac14(\I_4+\sum_{i=1}^3
t_i \sigma_i\otimes\sigma_i),
\end{eqnarray}
where $\sigma_i$ for $i=1,2,3$ are three Pauli operators as
$\sigma_1=\out{0}{1}+ \out{1}{0}$, $\sigma_2={\rm i}\out{0}{1}- {\rm
i}\out{1}{0}$, $\sigma_3=\proj{0}- \proj{1}$. Hence, a Bell-diagonal
state is specified by three real variables $t_1, t_2$, and $t_3$
such that
\begin{eqnarray}
\label{eq:dbell}
\begin{cases}
1-t_1-t_2-t_3\geqslant 0,\\
1-t_1+t_2+t_3\geqslant 0,\\
1+t_1-t_2+t_3\geqslant 0,\\
1+t_1+t_2-t_3\geqslant 0.
\end{cases}
\end{eqnarray}
Denote by $D_{\text{Bell}}$ the set of tuples $(t_1,t_2,t_3)$
satisfying the system of inequalities Eq. \eqref{eq:dbell}. Because
all the four eigenvalues of $\rho_{\text{Bell}}$ are in $[0,1]$, it
follows from Eq. \eqref{eq:defbelldiag} that $t_i\in[-1,1]$ for
$i=1,2,3$. That is, $D_{\text{Bell}}\subset[-1,1]^3$. Furthermore,
the Bell-diagonal states can be geometrically described by a
tetrahedron. One can show that a Bell-diagonal state is separable if
and only if $\abs{t_1}+\abs{t_2}+\abs{t_3}\leqslant 1$ holds.
Geometrically, the set of Bell-diagonal states is a tetrahedron and
the set of separable Bell-diagonal states is an octahedron
\cite{40Horodecki1996} denoted by $D_{\text{Bellsep}}$. By
calculation, all eigenvalues of $\rho^{\Gamma}_{\text{Bell}}$ are
\begin{eqnarray}
\label{eq:bellptm-1}
\begin{aligned}
x_1&=\frac{1+t_1-t_2-t_3}4,~~ x_2=\frac{1-t_1+t_2-t_3}4, \\
x_3&=\frac{1+t_1+t_2+t_3}4,~~ x_4=\frac{1-t_1-t_2+t_3}4.
\end{aligned}
\end{eqnarray}
It follows from Eq. \eqref{eq:pt-spectrum} that
\begin{eqnarray}
    \label{eq:pt23belldiag}
    \begin{aligned}
    p_2&=\frac{1+\sum^3_{i=1}t^2_i}4,\\
    p_3&=\frac{1+6t_1t_2t_3+3\sum^3_{i=1}t^2_i}{16}.
    \end{aligned}
\end{eqnarray}
Next, we set $p_2\in[\frac14,1]$, and use $p_2$ to bound $p_3$ from
above to below. It suffices to do the optimization:
\begin{eqnarray}
\label{eq:optdbell}
    \max_{(t_1,t_2,t_3)\in
D_{\text{Bell}}} ~~g(t_1,t_2,t_3)=t_1t_2t_3.
\end{eqnarray}
For the family of Bell diagonal states, we conclude
\begin{eqnarray}\label{eq:Bell}
\Phi^-_4(p_2)\leqslant p_3\leqslant \Phi^{\B}_4(p_2),
\end{eqnarray}
where $\Phi^{\B}_4(p_2)$ is piecewisely functioned as
\begin{eqnarray}
\Phi^{\B}_4(p_2) :=
\begin{cases}
\Phi^+_4(p_2),&\text{if }p_2\in[\frac14,\frac13],\\
\frac{(18p_2-1)+ 2\sqrt{2}(3p_2-1)^\frac32}{36},&\text{if }p_2\in[\frac13,\frac12],\\
\frac14,&\text{if }p_2\in[\frac12,1].\\
\end{cases}
\end{eqnarray}
This conclusion follows from the result of the optimization problem
in Eq. \eqref{eq:optdbell}. We put the optimizing process and the
proof of the inequality \eqref{eq:Bell} in Appendix~\ref{app:A2}. We
add the curve of function $\Phi^{\B}_4(p_2)$ in Fig~\ref{bdregion},
and then the light green colored region with both upper and lower
boundaries in Fig~\ref{bdregion} is corresponding to the family of
Bell-diagonal states from the inequality \eqref{eq:Bell}.
\begin{figure}[ht]\centering
{\begin{minipage}[b]{1\linewidth}
\includegraphics[width=1\textwidth]{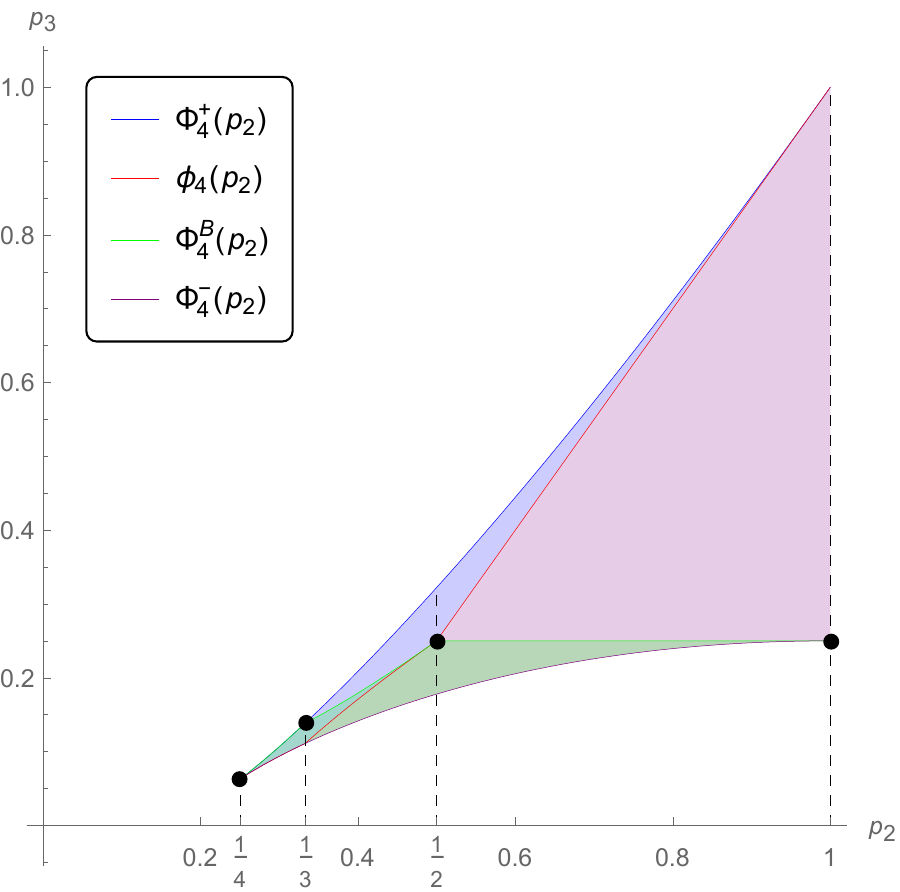}
\end{minipage}}
\caption{(Color online) The PT-moment pair $(p_2,p_3)$ for
Bell-diagonal state $\rho_{\text{Bell}}$ falls within the light
green colored region with both upper and lower boundaries; the
coordinates of these black points are (from left to right):
$\Pa{\frac14,\frac1{16}}, \Pa{\frac13,\frac5{36}},
\Pa{\frac12,\frac14},\Pa{1,\frac14}$.}\label{bdregion}
\end{figure}
Again, we formulate specific states corresponding to the border
lines of the family of Bell-diagonal states below.
\begin{itemize}
\item The curve $p_3=\tfrac14$ for $p_2\in[\tfrac12,1]$ corresponds to
\begin{eqnarray}
t\out{\phi^+}{\phi^+}+(1-t)\out{\phi^-}{\phi^-}
\end{eqnarray}
for $t\in[0,1]$.
\item The curve $p_3=\Phi^{\B}_4(p_2)$ for $p_2\in[\tfrac13,\tfrac12]$ corresponds to
\begin{eqnarray}
\frac14(\I_4+ t\sigma_1\otimes\sigma_1 +t \sigma_2\otimes\sigma_2 +
(1-2t)\sigma_3\otimes\sigma_3)
\end{eqnarray}
for $t\in[0,\frac13]$.
\item The curve $p_3=\Phi^{\B}_4(p_2)$ for $p_2\in[\tfrac14,\tfrac13]$ corresponds to
\begin{eqnarray}
\frac14(\I_4+ t \sum_i \sigma_i\otimes\sigma_i)
\end{eqnarray}
for $t\in[0,\frac13]$.
\end{itemize}

By calculation, we further obtain the fourth PT-moment
\begin{eqnarray}
p_4=\frac{1+24t_1t_2t_3+6\sum^3\limits_{i=1}t^2_i+\sum^3\limits_{i=1}t^4_i+6\sum\limits_{1\leqslant
i<j\leqslant3}t^2_i t^2_j}{64}.
\end{eqnarray}
We can also plot the region of the tuple $(p_2,p_3,p_4)$ for the
family of Bell-diagonal states. Moreover, if we consider the
negativity \cite{14Vidal2002} in place of $p_4$, which is formulated
by
\begin{eqnarray}
&&\cN(\rho_{\text{Bell}})=\tfrac{\norm{\rho_{\text{Bell}}^{\Gamma}}_1-1}{2}\\
&&=\tfrac18(\abs{1+t_1-t_2-t_3}+\abs{1-t_1+t_2-t_3}+\abs{1+t_1+t_2+t_3}+\abs{1-t_1-t_2+t_3}-4),\notag
\end{eqnarray}
we can plot the region of the tuple
$(p_2,p_3,\cN(\rho_{\text{Bell}}))$, see Fig~\ref{p2-p3-Neg}. The
negativity of Bell-diagonal states increases as $p_2$ or $p_3$
increases. Since $p_2$ and $p_3$ can be efficiently measured in
experiment, physically it means one can conveniently choose the
Bell-diagonal states with larger entanglement by $p_2$ or $p_3$.

Bell-diagonal entangled states are assisted maximally entangled
states \cite{44Zhao2021}, which can be decomposed as the convex
combination of maximally entangled states. Therefore, with the help
of the third party, Bell-diagonal entangled states can be
transformed into maximally entangled pure states by local
measurements and classical communication (LOCC). From the
perspective of entanglement distillation, Bell-diagonal entangled
states are resourceful and can widely be used in quantum information
processing tasks.
\end{exam}

\begin{figure}
    \centering
    \subfigure[The negativity of $\rho_{\rm Bell}$ versus the PT-moment pair $(p_2,p_3)$]{
    \begin{minipage}[b]{0.4\textwidth}
        \includegraphics[width=1\textwidth]{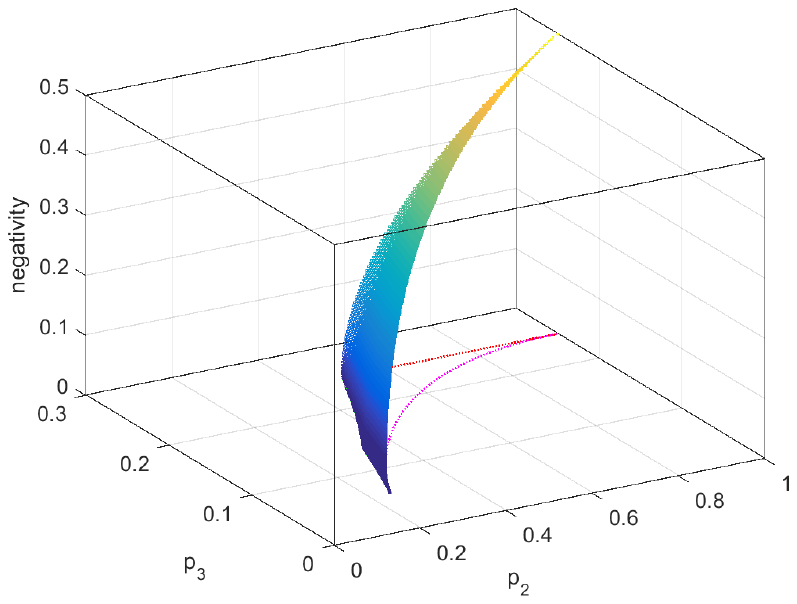}
    \end{minipage}
    \label{fig:p23neg}
    }\\
    \subfigure[The negativity of $\rho_{\rm Bell}$ versus the PT-moment $p_2$]{
    \begin{minipage}[h]{0.4\textwidth}
        \includegraphics[width=1\textwidth]{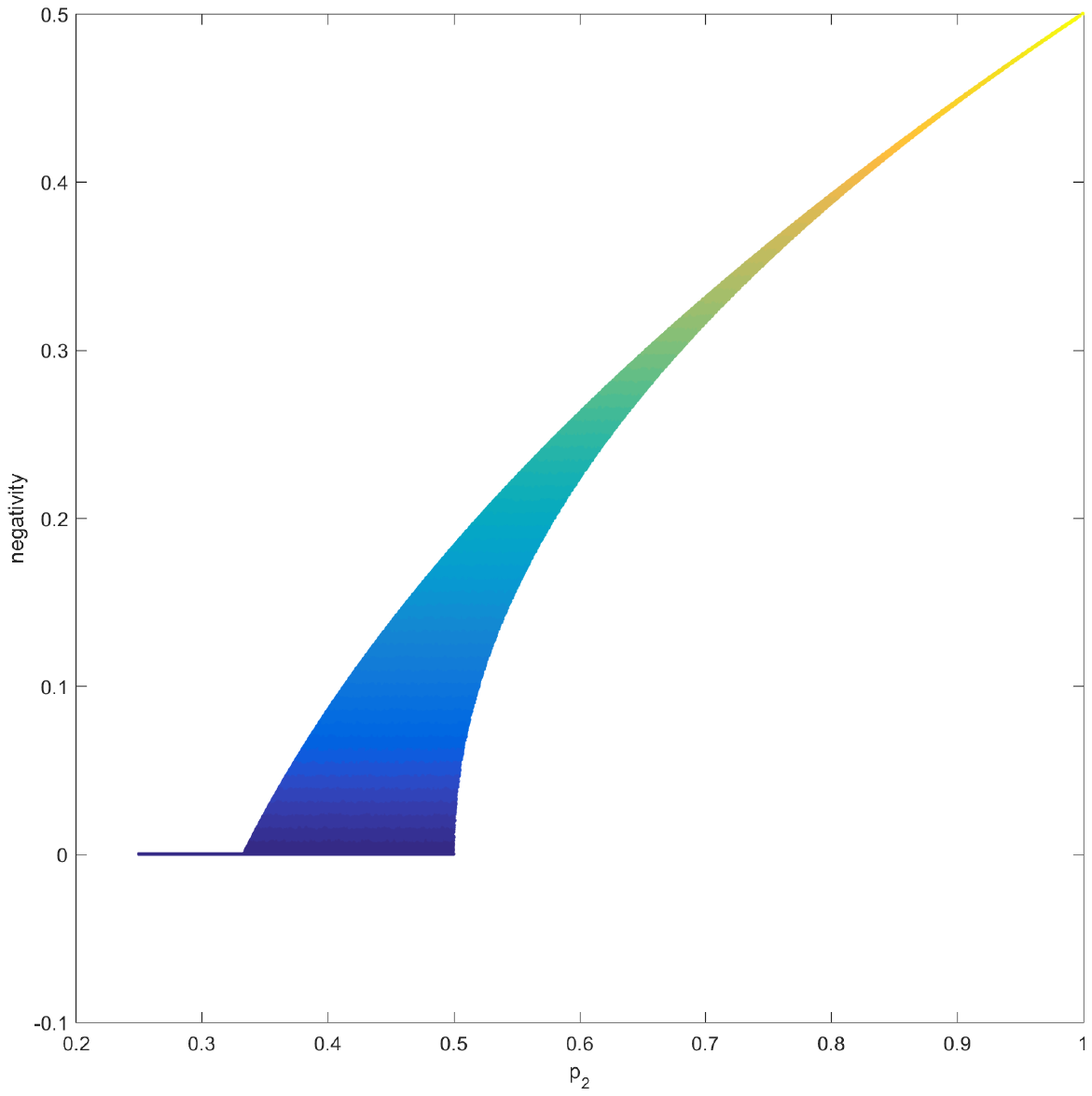}
    \end{minipage}
    \label{fig:p2neg}
    }
    \subfigure[The negativity of $\rho_{\rm Bell}$ versus the PT-moment $p_3$]{
    \begin{minipage}[h]{0.4\textwidth}
        \includegraphics[width=1\textwidth]{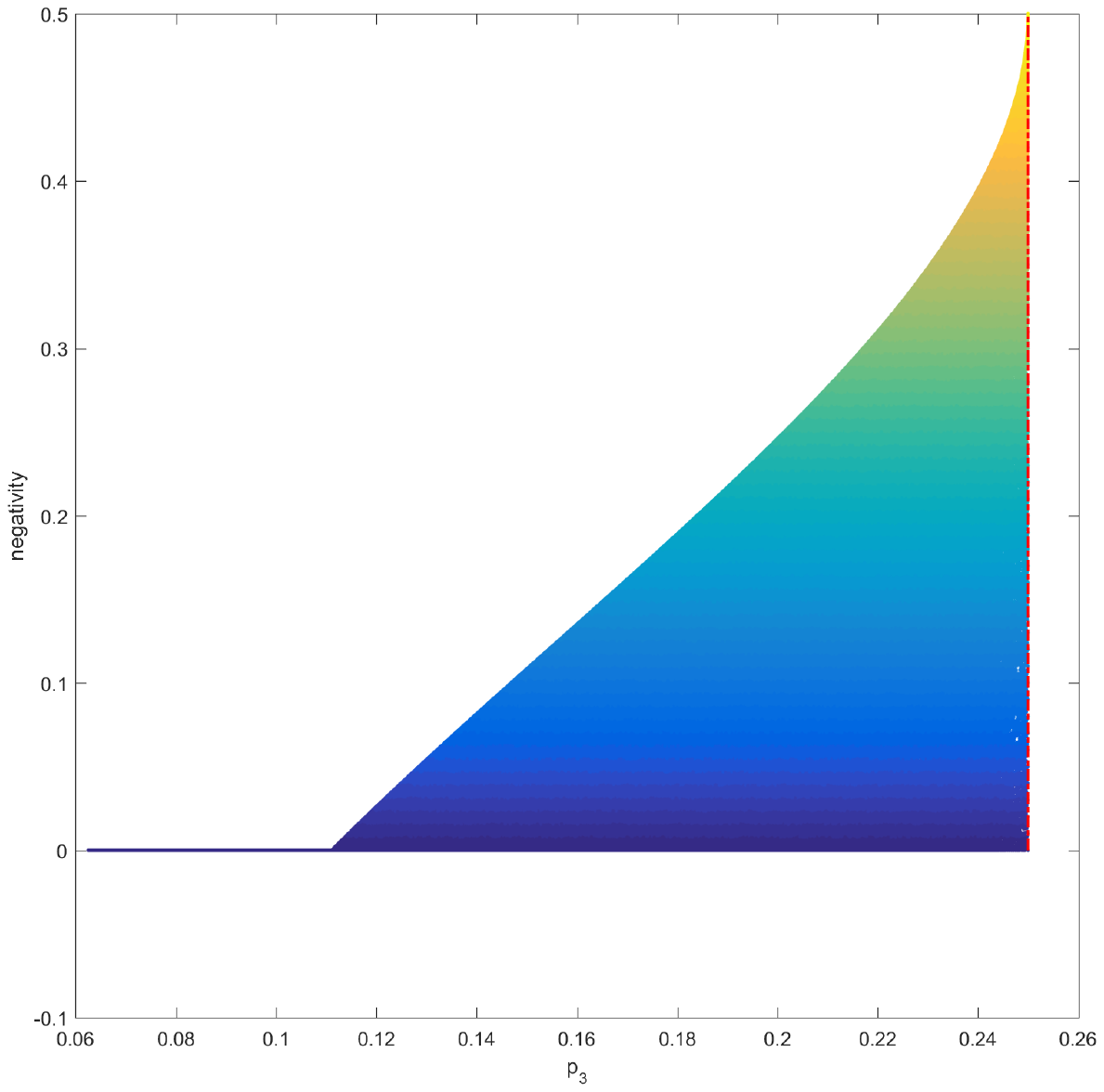}
    \end{minipage}
    }
    \caption{(Color Online) Here, $\rho_{\rm Bell}$ represents a two-qubit Bell-diagonal state, and the pairs $(p_2,p_3)$ are sampled from the light green colored region in Fig. \ref{bdregion}. The pairs $(p_2,p_3)$ on the left side of $\phi_4$-curve yield the negativity 0.}
    \label{p2-p3-Neg}
\end{figure}

\begin{exam}[The family of two-qubit X-states]
The $X$-state is a name for the class of mixed states whose density
matrix is in the X shape. Its quantum correlation such as quantum
discord \cite{45Luo2008,46Ali2010,47Maldonado2015}, one-way quantum
deficit \cite{48Ye2016,49Wang2015}, and quantum coherence including
coherence concurrence \cite{50Zhao2020} are investigated broadly.
Attributing to the symmetry of $X$-state formally, the study of its
quantumness seems to be more feasible than general mixed states. The
pure states that can be transformed into an $X$-state with
incoherent operations are also characterized in Ref.
\cite{51Wang2021}.

The so-called $X$-states in the two-qubit system are those states
whose density matrices are of the form:
\begin{eqnarray}
\rho_{\rX} = \Pa{\begin{array}{cccc}
                   \rho_{11} & 0 & 0 & \rho_{14} \\
                   0 & \rho_{22} & \rho_{23} & 0 \\
                   0 & \rho_{32} & \rho_{33} & 0 \\
                   \rho_{41} & 0 & 0 & \rho_{44}
                 \end{array}
}.
\end{eqnarray}
The $X$-state $\rho_{\rX}$ naturally satisfies the  unit trace and
the positivity conditions: (i) $\sum^4_{k=1}\rho_{kk}=1$ and (ii)
$\rho_{22}\rho_{33}\geqslant\abs{\rho_{23}}^2$ and
$\rho_{11}\rho_{44}\geqslant\abs{\rho_{14}}^2$. The
partial-transpose $\rho^{\Gamma}_{\rX}$ is given by
\begin{eqnarray}
\rho^{\Gamma}_{\rX} = \Pa{\begin{array}{cccc}
                   \rho_{11} & 0 & 0 & \rho_{23} \\
                   0 & \rho_{22} & \rho_{14} & 0 \\
                   0 & \rho_{41} & \rho_{33} & 0 \\
                   \rho_{32} & 0 & 0 & \rho_{44}
                 \end{array}
}.
\end{eqnarray}
Then the second and third PT-moments are, respectively,
\begin{eqnarray}
\label{eq:xstateptm}
\begin{aligned}
p_2&=\rho^2_{11}+\rho^2_{22}+\rho^2_{33}+\rho^2_{44}+2\abs{\rho_{23}}^2+2\abs{\rho_{14}}^2\\
p_3&=\rho^3_{11}+\rho^3_{22}+\rho^3_{33}+\rho_{44}^3+3(\rho_{11}+\rho_{44})\abs{\rho_{23}}^2+3(\rho_{22}+\rho_{33})\abs{\rho_{14}}^2.
\end{aligned}
\end{eqnarray}
By numerics from Eq. \eqref{eq:xstateptm}, we observe that the pairs
$(p_2, p_3)$ of PT-moments for two-qubit X-states fill the whole
region bounded by the $\Phi_4^\pm(p_2)$ curves, see
Fig~\ref{Xstate:p2-p3}. It implies that the family of two-qubit
$X$-states are almost the typical representative of all two-qubit
states when considering PT-moments.
\begin{figure}[ht]\centering
{\begin{minipage}[b]{0.6\textwidth}
\includegraphics[width=1\textwidth]{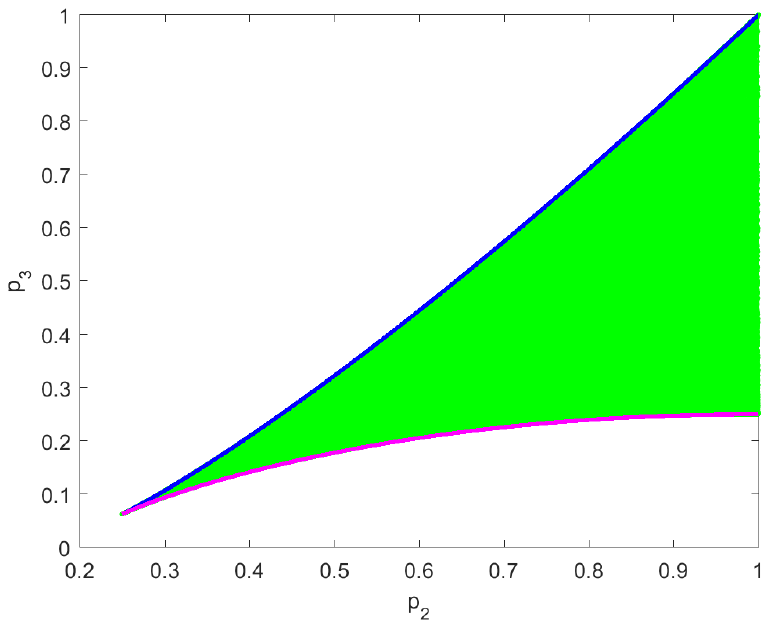}
\end{minipage}}
\caption{(Color Online) The PT-moment pair $(p_2, p_3)$ for X-state
$\rho_{\rX}$ falls within the green colored region. The upper blue
curve and the lower magenta curve correspond to $\Phi_4^\pm(p_2)$,
respectively, where $p_2 \in [\frac14,1]$. }\label{Xstate:p2-p3}
\end{figure}
\end{exam}

\subsection{multipartite system}\label{subsec:multi}

In this subsection, we keep applying our results in Sec.
\ref{sect:mainresult} to further detect the entanglement of
high-dimensional and multipartite states. Firstly, we consider a
bipartite state $\rho$ supported on $\bbC^{2m}\otimes\bbC^{2n}$ with
positive integers $m$ and $n$. We regard the system of $\rho$
consisted of four particles $A,B,C$ and $D$, where the dimensions of
the four subsystems are $2,m,2,n$ respectively. By tracing out the
subsystems $B$ and $D$, the reduced density operator $\rho_{AC}$ is
a two-qubit state. Then we can experimentally detect the
entanglement of $\rho_{AC}$ by measuring its PT-moments and using
Theorem \ref{th:A1}. If $\rho_{AC}$ is entangled then one can verify
the  entanglement of bipartite state $\rho_{AB:CD}$. Using a similar
idea, one can detect the entanglement of a multiqubit or
multipartite state.

For example, we consider the $n$-qubit mixed state which is the
mixture of $n$-qubit GHZ and W states, namely
\begin{eqnarray}\label{eq:nqubitghzw-def}
\rho^{(n)}= \lambda\proj{\rG\rH\rZ_n} +(1-\lambda)\proj{\rW_n},
\end{eqnarray}
where $\lambda\in[0,1]$, and
\begin{eqnarray}\label{eq:ghzandw}
\begin{aligned}
\ket{\rG\rH\rZ_n}&=\tfrac{1}{\sqrt2}(\ket{0}^{\otimes
n}+\ket{1}^{\otimes n}),\\
\ket{\rW_n}&=\tfrac{1}{\sqrt
n}(\ket{0\cdots01}+\ket{0\cdots10}+...+\ket{10\cdots0}).
\end{aligned}
\end{eqnarray}
Since $\rho^{(n)}$ is invariant under any permutation of the $n$
subsystems, by computing we obtain any bipartite reduced density
operator of $\rho^{(n)}$ is
\begin{eqnarray}\label{eq:nqubitrdm}
\rho^{(n)}_{AB}=
\tfrac{\lambda}{2}(\proj{00}+\proj{11})+\frac{1-\lambda}{n}(n-2)\proj{00}+\tfrac{1-\lambda}{n}(\ket{01}+\ket{10})(\bra{01}+\bra{10}).
\end{eqnarray}
By straightforward computation, we obtain the four eigenvalues of
its partial transpose are
\begin{eqnarray}    \label{eq:nqubitrdmeig-1}
\begin{aligned}
    &\tfrac{1-\lambda}n,\quad \tfrac{1-\lambda}n,\\
    &\tfrac{1}{2}+\tfrac{(1-\lambda)(\sqrt{\delta_n}-2)}{2 n}, \\
    &\tfrac{1}{2}-\tfrac{(1-\lambda)(\sqrt{\delta_n}+2)}{2 n},
\end{aligned}
\end{eqnarray}
where $\delta_n=n^2-4n+8\geqslant 4$. One can verify that the first
three eigenvalues are always non-negative for any $n\geqslant 2$
when $\lambda\in[0,1]$. Hence, it follows from the PPT criterion
that the two-qubit state $\rho^{(n)}_{AB}$ is entangled if and only
if the last eigenvalue in Eq. \eqref{eq:nqubitrdmeig-1} is negative.
Specifically, by direct calculation we obtain that
\begin{itemize}
\item If $n=2$, then $0\leqslant \lambda<\frac12$ corresponds to
$\rho_{AB}$ being entangled;
\item If $n>2$, then $0\leqslant \lambda<\frac{\delta_n-n\sqrt{\delta_n}+2n-4}{(n-2)^2}$ corresponds to
$\rho_{AB}$ being entangled.
\end{itemize}
Note that
$\lim\limits_{n\to\infty}\frac{\delta_n-n\sqrt{\delta_n}+2n-4}{(n-2)^2}=0$.
It follows from Eq. \eqref{eq:nqubitrdm} that entangled
$\rho^{(n)}_{AB}$ asymptotically approaches to
\begin{eqnarray}
\tfrac{n-2}{n}\proj{00}+\tfrac{1}{n}(\ket{01}+\ket{10})(\bra{01}+\bra{10})
\end{eqnarray}
as $n\to+\infty$. Furthermore, we can also conclude that the
original $n$-qubit state $\rho^{(n)}$ in Eq.
\eqref{eq:nqubitghzw-def} which generates entangled two-qubit
reduced density operators asymptotically approaches to
$\proj{\rW_n}$ as $n\to+\infty$.

\begin{figure}[ht]\centering
{\begin{minipage}[b]{1\linewidth}
\includegraphics[width=1\textwidth]{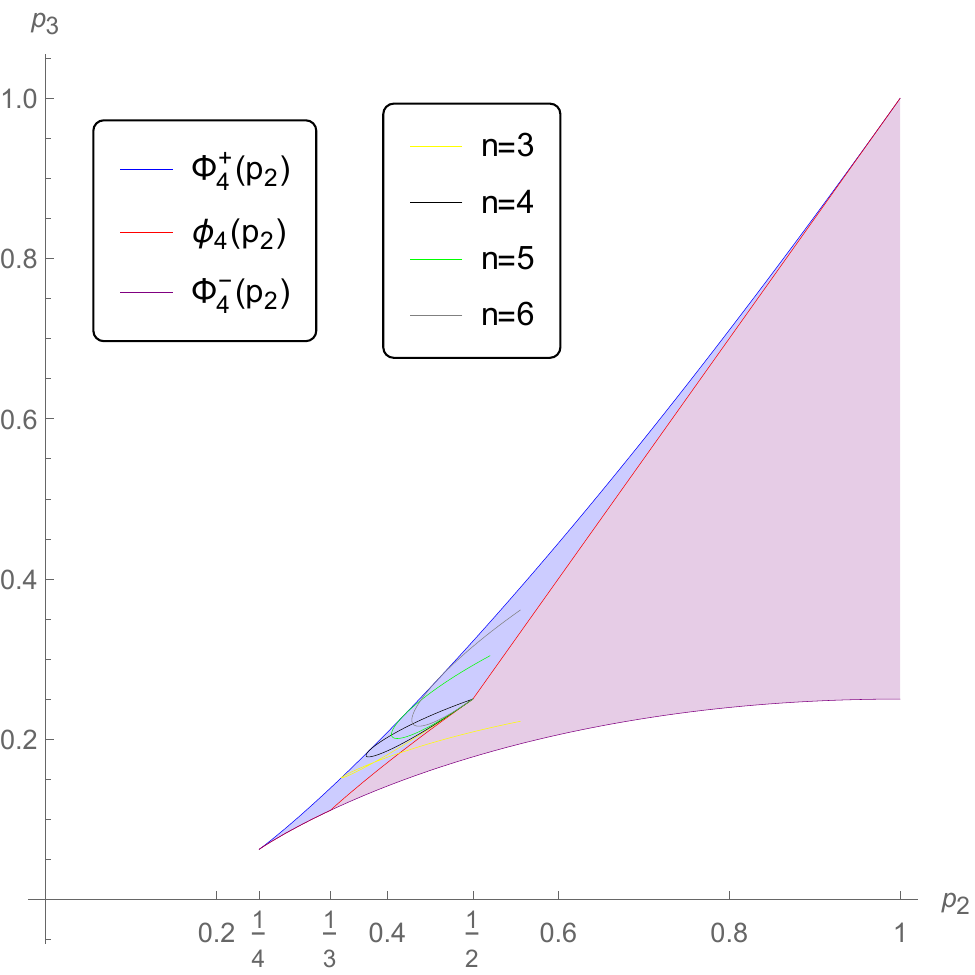}
\end{minipage}}
\caption{(Color Online) The PT-moment pair $\Pa{p^{(n)}_2,
p^{(n)}_3}$ of two-qubit state $\rho^{(n)}_{AB}$ in Eq.
\eqref{eq:nqubitrdm} for $n=3,4,5,6$.}\label{nqubit}
\end{figure}

Based on the four eigenvalues formulated in Eq.
\eqref{eq:nqubitrdmeig-1}, we further calculate the PT-moments $p_2$
and $p_3$ as follows.
\begin{eqnarray}
\begin{aligned}
p_2(\lambda)=&\frac{h_{4,16}\lambda^2 - 2h_{6,16}\lambda + 2h_{4,8}}{2n^2},\\
p_3(\lambda)=&\frac{6h_{4,8}\lambda^3+3(n-4)h_{6,12}\lambda^2-6
(n-2) h_{6,12}\lambda+4(nh_{6,15}-12)}{4n^3},
\end{aligned}
\end{eqnarray}
where $h_{x,y}:=n^2-x n+ y$. We investigate the pairs of PT-moments
$\Pa{p^{(n)}_2(\lambda),p^{(n)}_3(\lambda)}$, where
$\lambda\in[0,1]$, similar to that in Sec. \ref{subsec:twoqubit}. We
plot the parametric curves for $n=3,4,5,6$ in Fig.~\ref{nqubit}.
From Fig.~\ref{nqubit} we can see in which interval of $p_2$ the
two-qubit reduced state $\rho^{(n)}_{AB}$ falls within the region of
separable states defined by Theorem \ref{th:A1}, and the same for
the region of entangled states. Then for the interval of $p_2$
corresponding to entangled $\rho^{(n)}_{AB}$, we may experimentally
verify the entanglement of the original $n$-qubit state $\rho^{(n)}$
in Eq. \eqref{eq:nqubitghzw-def} because the PT-moments are
measurable. Furthermore, since $\rho^{(n)}$ is a symmetric state, it
is either genuinely entangled or fully separable. This means we
present an operational way of verifying multiqubit genuine
entanglement using Theorem \ref{th:A1}.

\section{Concluding remarks}\label{sec:con}

In this short report, we derived an experimentally efficient
separability criterion of two-qubit states based on the measurable
PT-moments. Specifically, we concluded that a two-qubit state
$\rho_{AB}$ is entangled if and only if the pair of its PT-moments
$(p_2,p_3)$ belonging to the set $R_{\mathrm{ent}}$ defined by Eq.
\eqref{eq:defentsepset} following from Theorem \ref{th:A1}. We also
demonstrated our main result Theorem \ref{th:A1} geometrically in
Fig~\ref{ptregion}. Our main contribution was determining the lowest
curve $\Phi^-_4(p_2)$ of $R_{\mathrm{ent}}$ in Fig~\ref{ptregion}.
Then, in Sec. \ref{subsec:twoqubit}, we applied our main results to
several widely-used families of two-qubit states such as the
families of Werner states and Bell-diagonal states. We analytically
identified the regions corresponding to the entangled states in
these families. Moreover, we also extended our study on the
two-qubit system to high-dimensional and multipartite systems in Sec
\ref{subsec:multi}. We mainly characterized the entanglement in the
$n$-qubit state mixed by $n$-qubit GHZ and W states. Through
identifying the region $R_{\mathrm{ent}}$ of the reduced two-qubit
state in Fig. \ref{nqubit}, we can experimentally verify the genuine
entanglement of such multiqubit states. We hope the results obtained
in this work can shed new light on some related problems in quantum
information theory, such as to quantify both the values and numbers
of the negative eigenvalues in the spectrum of $\rho_{AB}^{\Gamma}$.

\subsection*{Acknowledgments}
The authors would like to thank the anonymous reviewers and the
editor for their detailed and valuable comments which are very
helpful to improve the standard of this paper. This work is
supported by the NSF of China under Grant Nos. 11971140, 12171044
and 11871089, also supported by the National Key Research and
Development Program of China under grant No. 2021YFA1000600.


\newpage
\appendix
\appendixpage
\addappheadtotoc

\section{The proof of Theorem~\ref{th:A1}}\label{app:A1}

Consider the objective function
$$
f(x,y,z)=x^3+y^3+z^3+(1-x-y-z)^3\equiv p_3,\quad (x,y,z)\in[0,1]^3
$$
subjective to the constraints
$$
x+y+z=s,\quad x^2+y^2+z^2+(1-x-y-z)^2=t\equiv p_2
$$
where both parameters $s\in[1,\tfrac32]$ and $t\in[\tfrac14,1]$ are
fixed. Clearly $t-(1-s)^2\geqslant0$ for $s\in[1,\tfrac32]$ and
$t\in[\tfrac14,1]$. Recall that
\begin{eqnarray*}
x^3+y^3+z^3-3xyz &=& (x+y+z)(x^2+y^2+z^2-xy-yz-zx)\\
&=&s\Br{t-(1-s)^2-\frac12(s^2+(1-s)^2-t)}\\
&=&\frac{1}{2} s \Pa{-4 s^2+6 s+3 t-3}\geqslant0.
\end{eqnarray*}
Thus our objective function is reduced as
\begin{eqnarray*}
f(x,y,z) = 3xyz+\frac32s(-2 s^2+4 s+t-3)+1.
\end{eqnarray*}
Denote by $g(x,y,z)=xyz$, where $(x,y,z)\in[0,1]^3$. Then
$$
\min f(x,y,z) = 3\min g(x,y,z) + \frac32s(-2 s^2+4 s+t-3)+1
$$
for given $s$ and $t$. Denote $M(t,s):=f_{\max}$ and
$m(t,s):=f_{\min}$. Our problem becomes to consider the optimization
of the function $g(x,y,z)$ subjective to the constraints:
$$
x+y+z=s,\quad x^2+y^2+z^2=t-(1-s)^2.
$$
Apparently the plain $x+y+z=s$ must intersect with the sphere
$x^2+y^2+z^2=t-(1-s)^2$, which means that
\begin{eqnarray*}
\frac{\abs{0+0+0-s}}{\sqrt{1^2+1^2+1^2}}\leqslant
\sqrt{t-(1-s)^2}\Longleftrightarrow \frac13< t\leqslant 1,
1<s\leqslant\frac{3+\sqrt{3(4t-1)}}4\Pa{\leqslant\frac32}.
\end{eqnarray*}
The plain $x+y+z=s$ is tangent to the sphere $x^2+y^2+z^2=t-(1-s)^2$
iff $s=\frac{3+\sqrt{3(4t-1)}}4$, where $t\in[\tfrac13,1]$. At this
time,
\begin{eqnarray*}
x=y=z=\frac s3= \frac{3+\sqrt{3(4t-1)}}{12}.
\end{eqnarray*}
Then
\begin{eqnarray*}
f(s/3,s/3,s/3)&=&3g(s/3,s/3,s/3)+\frac{3(2-t)-2\sqrt{3(4t-1)}}{16}
\\
&=&\frac1{24} \Br{3(6t-1)- \sqrt{3}(4t-1)^{\frac32}}\quad
(t\in[\tfrac13,1]).
\end{eqnarray*}
(i) If moreover $s>\sqrt{2}\sqrt{t-(1-s)^2}$ as well, i.e.,
$$
\frac{s^2}3\leqslant t-(1-s)^2<\frac{s^2}2
$$
that is, $(t,s)\in R_1\cup R_2$ where
\begin{eqnarray*}
R_1&=&\Set{(t,s): \frac13\leqslant t<\frac12, 1\leqslant s\leqslant
\frac{3+\sqrt{3(4t-1)}}4},\\
R_2&=&\Set{(t,s):\frac12\leqslant
t\leqslant1,\frac{2+\sqrt{2(3t-1)}}3<s\leqslant
\frac{3+\sqrt{3(4t-1)}}4},
\end{eqnarray*}
the plane $x+y+z=s$ must intersection with the sphere
$x^2+y^2+z^2=t-(1-s)^2$ in a whole circle of the first octant;
moreover this circle does not touch the three coordinate plains. By
using Lagrange's multiplier method, we let Lagrange's function be
$$
G(x,y,z,\mu,\nu) = xyz+\mu(x+y+z-s)+\nu(x^2+y^2+z^2-t+(1-s)^2).
$$
Then
\begin{eqnarray*}
\begin{cases}
\frac{\partial G}{\partial x} &= \mu +2 \nu x+y z=0\\
\frac{\partial G}{\partial y} &= \mu +2 \nu y+x z=0\\
\frac{\partial G}{\partial z} &= \mu +2 \nu z+x y=0\\
\frac{\partial G}{\partial \mu}&=x+y+z-s=0\\
\frac{\partial G}{\partial \nu}&=x^2+y^2+z^2-t+(1-s)^2=0
\end{cases}
\end{eqnarray*}
Based on the previous three equations, we get that
\begin{eqnarray*}
\begin{cases}
y z&= -\mu -2 \nu x\\
x z&= -\mu -2 \nu y\\
x y&= -\mu -2 \nu z\\
\end{cases}\Longrightarrow \begin{cases}
x y z&= -x(\mu +2 \nu x)\\
x y z&= -y(\mu +2 \nu y)\\
x y z&= -z(\mu +2 \nu z)\\
\end{cases}
\end{eqnarray*}
This implies that
$$
2\nu x^2+\mu x =2\nu y^2+\mu y =2\nu z^2+\mu z.
$$
Let $\varphi(r)=2\nu r^2+\mu r$ be a polynomial of 2nd degree in the
argument $r$. Then the above fact means that
$\varphi(x)=\varphi(y)=\varphi(z)$, i.e., $\varphi$ attains the same
value at three points $x,y,z$. This indicates that at least two
points of three points $x,y,z$ must be equal. Due to the symmetry of
all permutations of $x,y,z$ for the objective function and
constraints, without loss of generality, we assume that $x=y$. Thus
\begin{eqnarray*}
\begin{cases}
x=2\nu\\
y=2\nu\\
z=s-4\nu\\
\mu = 2\nu(2\nu-s)
\end{cases}
\end{eqnarray*}
Therefore we get that
\begin{eqnarray*}
\nu_i = \frac{2s+(-1)^i\sqrt{2(-4 s^2+6 s+3 t-3)}}{12}\quad(i=1,2).
\end{eqnarray*}
is the only two solutions allowing $(x,y,z)$ in the first orthant.
Finally, we get two extremal points $(x_i,y_i,z_i)(i=1,2)$, where
\begin{eqnarray*}
\begin{cases}
x_1 =& \frac{2s-\sqrt{2(-4 s^2+6 s+3 t-3)}}6\\
y_1 =& \frac{2s-\sqrt{2(-4 s^2+6 s+3 t-3)}}6\\
z_1 =& \frac{s+\sqrt{2(-4 s^2+6 s+3 t-3)}}3
\end{cases}\quad\text{and}\quad
\begin{cases}
x_2 =& \frac{2s+\sqrt{2(-4 s^2+6 s+3 t-3)}}6\\
y_2 =& \frac{2s+\sqrt{2(-4 s^2+6 s+3 t-3)}}6\\
z_2 =& \frac{s-\sqrt{2(-4 s^2+6 s+3 t-3)}}3
\end{cases}
\end{eqnarray*}
Note that $(x_1,y_1,z_1)\in[0,1]^3$ if and only if $\frac13\leqslant
t\leqslant 1$ and $1\leqslant s\leqslant\frac{3+\sqrt{3(4t-1)}}4$;
$(x_2,y_2,z_2)\in(0,1)^3$ if and only if $(t,s)\in R_1\cup R_2$. In
a word, when $(t,s)\in R_1\cup R_2$, we have two extremal points
$(x_i,y_i,z_i)(i=1,2)$. Thus $g_i=g(x_i,y_i,z_i)$. The optimal
values of $f$ are given by $f_i=x^3_i+y^3_i+z^3_i+(1-s)^3$,
respectively, i.e.,
\begin{eqnarray*}
\begin{cases}
M(t,s)=f_1=&\frac{1}{18} \Pa{-40 s^3+90 s^2+18 s t-72 s+18+\sqrt{2}
\Pa{-4
s^2+6 s+3 t-3}^{\frac32}}\\
m(t,s)=f_2=&\frac{1}{18} \Pa{-40 s^3+90 s^2+18 s t-72 s+18-\sqrt{2}
\Pa{-4 s^2+6 s+3 t-3}^{\frac32}}
\end{cases}
\quad(\forall (t,s)\in R_1\cup R_2).
\end{eqnarray*}
(ii) If $s\leqslant\sqrt{2}\sqrt{t-(1-s)^2}$, i.e.,
$$
t-(1-s)^2\geqslant\frac{s^2}2\Pa{\geqslant \frac{s^2}3}
$$
that is, $(t,s)\in R_3=\Set{(t,s): \frac12<t\leqslant1, 1\leqslant
s\leqslant \frac{2+\sqrt{2(3t-1)}}3}$, the plane $x+y+z=s$ must
intersection with the sphere $x^2+y^2+z^2=t-(1-s)^2$ in three arcs
separately of the first orthant; moreover such arcs touch the
coordinate plains separately. Thus $g_{\min}=0$. Thus
$f_{\min}=g_{\min}+\frac32s(-2 s^2+4 s+t-3)+1=\frac32s(-2 s^2+4
s+t-3)+1$. That is,
$$
m(t,s)=\frac32s(-2 s^2+4 s+t-3)+1\quad (\forall(t,s)\in R_3).
$$
Using the analogous method as above, we get that
$x=y=2\nu,z=s-4\nu$, where
$\nu=\frac{2s-\sqrt{2(-4s^2+6s+3t-3)}}{12}$. Thus the maximal value
of $f$ is still of the form
\begin{eqnarray*}
M(t,s) = \frac{1}{18} \Pa{-40 s^3+90 s^2+18 s t-72 s+18+\sqrt{2}
\Pa{-4 s^2+6 s+3 t-3}^{\frac32}}\quad (\forall(t,s)\in R_3).
\end{eqnarray*}
In summary, we get that $p_3\geqslant m(t,s)$ where $(t,s)\in
R:=R_1\cup R_2\cup R_3$, where
\begin{eqnarray*}
m(t,s) =
\begin{cases}
\frac{1}{18} \Br{-40 s^3+90 s^2+18 s t-72 s+18-\sqrt{2} \Pa{-4 s^2+6
s+3 t-3}^{\frac32}},&\text{if } \forall (t,s)\in R_1\cup R_2;\\
\frac32s(-2 s^2+4 s+t-3)+1,&\text{if } \forall(t,s)\in R_3.
\end{cases}
\end{eqnarray*}
In order to eliminate $s$, we make the optimization over the section
for fixed $t$:
\begin{eqnarray*}
\max_{s\in R_t}M(t,s)\geqslant f(x,y,z)=p_3\geqslant\min_{s\in
R_t}M(t,s)\\
\max_{s\in R_t}m(t,s)\geqslant f(x,y,z)=p_3\geqslant\min_{s\in
R_t}m(t,s)
\end{eqnarray*}
Note that
\begin{eqnarray*}
\min_{s\in R_t}M(t,s) =
M\Pa{t,\frac{3+\sqrt{3(4t-1)}}4} =\frac{3(6t-1)- \sqrt{3}(4t-1)^{\frac32}}{24}\\
\min_{s\in R_t}m(t,s) =m\Pa{t,\frac{3+\sqrt{3(4t-1)}}4}=
\frac{3(6t-1)- \sqrt{3}(4t-1)^{\frac32}}{24};
\end{eqnarray*}
which means this value is a global minimal value, can be attained
when $x=y=z$, as previously shown. We also see that
\begin{eqnarray*}
\max_{s\in R_t}M(t,s) = M(t,1)
=\frac{2(9t-2)+\sqrt{2}(3t-1)^{\frac32}}{18}
\end{eqnarray*}
and
\begin{eqnarray*}
\max_{s\in (R_1\cup R_2)_t}m(t,s) &=&\frac{2(9t-2)-\sqrt{2}(3t-1)^{\frac32}}{18},\quad \frac13<t<\frac12\\
 \max_{s\in
(R_3)_t}m(t,s)&=&m(t,1) = \frac{3t-1}2,\quad \frac12\leqslant t<1
\end{eqnarray*}
that is,
\begin{eqnarray*}
\max_{s\in R_t}m(t,s)
&=&\max\Set{\frac{2(9t-2)-\sqrt{2}(3t-1)^{\frac32}}{18},\frac{3t-1}2},\quad
\quad \frac13\leqslant t\leqslant 1.
\end{eqnarray*}
Therefore
\begin{eqnarray*}
\min\Set{\max_{s\in R_t}M(t,s),\max_{s\in R_t}m(t,s)}\geqslant
f(x,y,z)=p_3\geqslant\max\Set{\min_{s\in R_t}M(t,s),\min_{s\in
R_t}m(t,s)}.
\end{eqnarray*}
Here
\begin{eqnarray*}
\min\Set{\max_{s\in R_t}M(t,s),\max_{s\in
R_t}m(t,s)}&=&\max\Set{\frac{2(9t-2)-\sqrt{2}(3t-1)^{\frac32}}{18},\frac{3t-1}2},\quad\frac13\leqslant
t\leqslant 1\\
\max\Set{\min_{s\in R_t}M(t,s),\min_{s\in R_t}m(t,s)}
&=&\frac{3(6t-1)-
\sqrt{3}(4t-1)^{\frac32}}{24},\quad\frac13\leqslant t\leqslant 1.
\end{eqnarray*}
Now rewrite the above inequality by replacing $t$ as $p_2$. We get
the desired:$\phi_4(p_2)\geqslant p_3\geqslant \Phi^-_4(p_2)$ for
$p_2\in[\tfrac13,1]$. Due to the fact that $p_3=\phi_4(p_2)$ for
$p_2\in[\tfrac13,1]$ is compatible with $\rho_{AB}$ being separable,
we finally obtain that
\begin{eqnarray*}
\phi_4(p_2)>p_3\geqslant \Phi^-_4(p_2),\quad p_2\in(\tfrac13,1].
\end{eqnarray*}

\section{The proof of inequality~\eqref{eq:Bell}}\label{app:A2}

In fact, we note that the family of Werner states is a subset of the
family of Bell-diagonal states. This indicates that
\begin{eqnarray*}
p_3\geqslant \phi^-_4(p_2).
\end{eqnarray*}
This inequality is saturated for Werner states. Next, we show that
$p_3\leqslant \Phi^{\rB}_4(p_2)$, where $p_2\in[\frac14,1]$. \\
(i) Assume that $p_2\in[\frac12,1]$ is fixed. Next we show that
$p_3\leqslant\frac14$. Recall that
\begin{eqnarray*}
p_3=\frac{1+3\sum^3_{i=1}t_i+6t_1t_2t_3}{16} =
\frac{(6p_2-1)+3t_1t_2t_3}8.
\end{eqnarray*}
Thus it suffices to show that $t_1t_2t_3\leqslant
1-2p_2(\leqslant0)$ when $(t_1,t_2,t_3)\in D_{\text{Bell}}$ and
$p_2\in[\frac12,1]$. This amounts to do the following optimization
problem: For $g(t_1,t_2,t_3)=t_1t_2t_3$,
\begin{eqnarray*}
&~~~~~~~\max& g(t_1,t_2,t_3)\\
&\text{subject to: }& (t_1,t_2,t_3)\in D_{\text{Bell}}\\
&&t^2_1+t^2_2+t^2_3=4p_2-1\in[1,3].
\end{eqnarray*}
Note that the sphere (which is denoted by $S$)
$t^2_1+t^2_2+t^2_3=4p_2-1\in[1,3]$ intersect with the 3D region
$D_{\text{Bell}}$ at four congruent pieces (outside of
$D_{\text{Bellsep}}$), respectively, in 2nd, 4th, 5th, 7th orthant.
See the following Figure~\ref{bellop2}.
\begin{figure}[ht]\centering
{\begin{minipage}[b]{0.5\linewidth}
\includegraphics[width=1\textwidth]{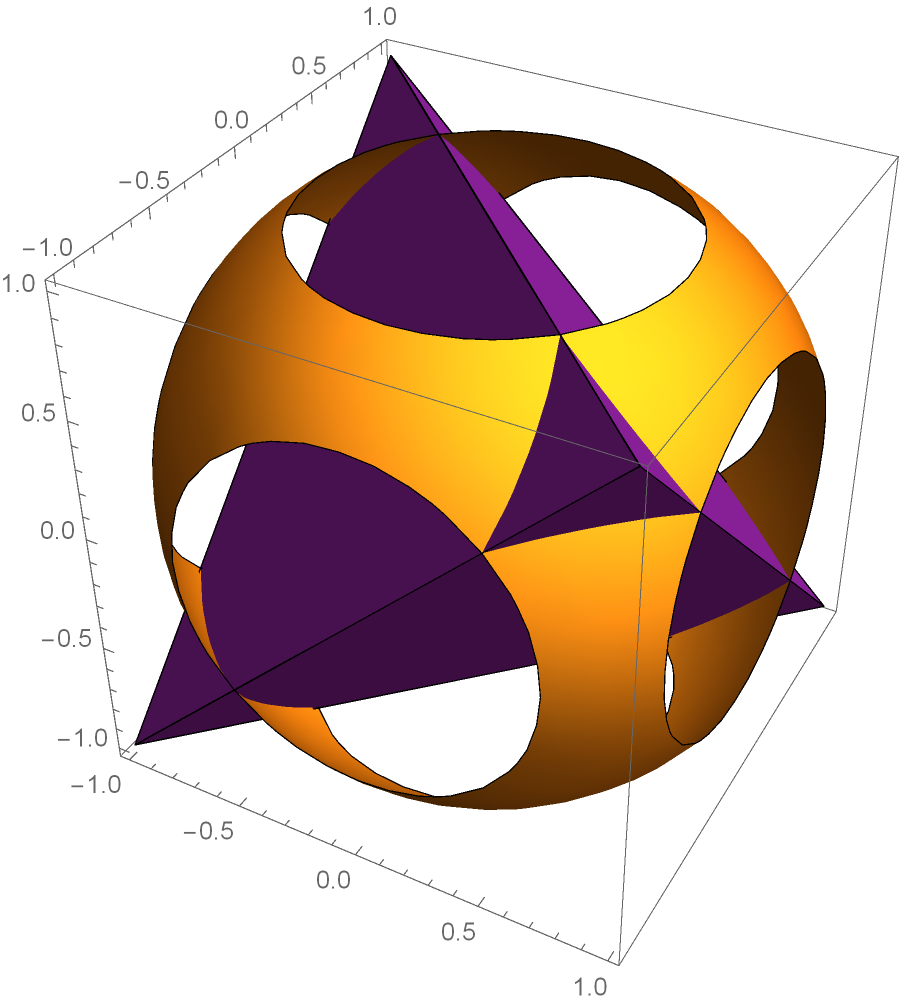}
\end{minipage}}
\caption{The optimization problem for Bell-diagonal states in the
case where $p_2\in[\frac12,1]$.}\label{bellop1}
\end{figure}
For instance, we focus the pieces in the 4th orthant. When
$t_1=1,t_2=-\sqrt{2p_2-1},t_3=\sqrt{2p_2-1}$,
$g(t_1,t_2,t_3)=1-2p_2$ attain its maximal value.\\
(ii) Assume that $p_2\in[\frac13,\frac12]$ is fixed. This amounts to
do the following optimization problem: For
$g(t_1,t_2,t_3)=t_1t_2t_3$,
\begin{eqnarray*}
&~~~~~~~\max& g(t_1,t_2,t_3)\\
&\text{subject to: }& (t_1,t_2,t_3)\in D_{\text{Bell}}\\
&&t^2_1+t^2_2+t^2_3=4p_2-1\in[\tfrac13,1].
\end{eqnarray*}
Note that the sphere $S: t^2_1+t^2_2+t^2_3=4p_2-1\in[\tfrac13,1]$
intersect with $D_{\text{Bellsep}}$ at nontrivial nonempty set. As
already obtained, $p_3$ will attain more larger values on separable
states (corresponding to those points in $D_{\text{Bellsep}}$)
versus entangled states (corresponding to those points in
$D_{\text{Bell}}\backslash D_{\text{Bellsep}}$) when the purity
$p_2$ is fixed. It suffices to consider the optimization problem:
\begin{eqnarray*}
&~~~~~~~\max& g(t_1,t_2,t_3)\\
&\text{subject to: }& \abs{t_1}+\abs{t_2}+\abs{t_3}=1,\text{ i.e., } (t_1,t_2,t_3)\in\partial D_{\text{Bellsep}}\\
&&t^2_1+t^2_2+t^2_3=4p_2-1\in[\tfrac13,1].
\end{eqnarray*}
In fact, by symmetry, we can further specialize the above boundary
condition $(t_1,t_2,t_3)\in\partial D_{\text{Bellsep}}$ to the case
$t_1+t_2+t_3=1$, where $t_i\geqslant0$ for $i=1,2,3$. Thus, for
instance, $g(t_1,t_2,t_3)$ attains its maximal value
$\frac{(18p_2-1)+2\sqrt{2}(3p_2-1)^{\frac32}}{36}$ when
$t_1=t_2=\frac{1-\sqrt{2(3p_2-1)}}3$ and
$t_3=\frac{1+2\sqrt{2(3p_2-1)}}3$, where
$p_2\in[\frac13,\frac12]$.\\
See the following Figure~\ref{bellop2}.
\begin{figure}[ht]\centering
{\begin{minipage}[b]{0.5\linewidth}
\includegraphics[width=1\textwidth]{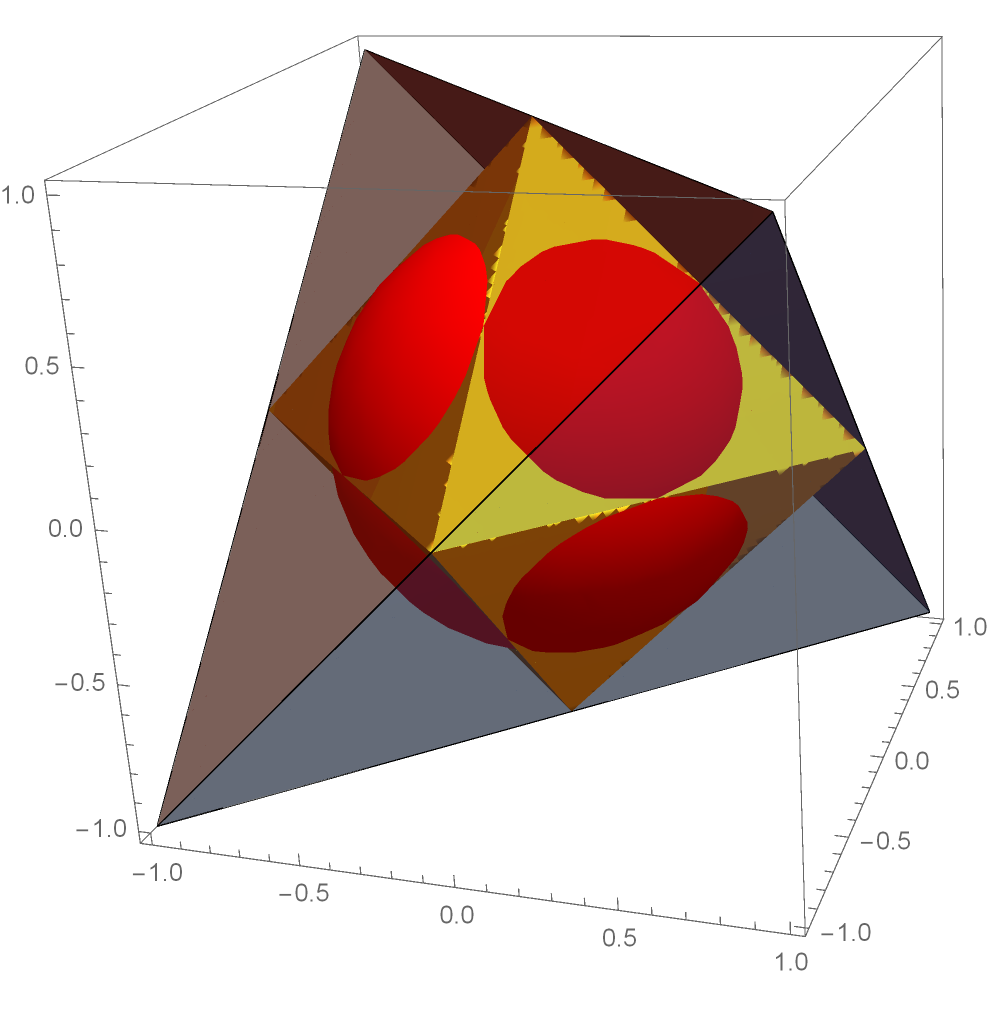}
\end{minipage}}
\caption{The optimization problem for Bell-diagonal states in the
case where $p_2\in[\frac13,\frac12]$.}\label{bellop2}
\end{figure}
(iii) Assume that $p_2\in[\frac14,\frac13]$ is fixed. This amounts
to do the following optimization problem: For
$g(t_1,t_2,t_3)=t_1t_2t_3$,
\begin{eqnarray*}
&~~~~~~~\max& g(t_1,t_2,t_3)\\
&\text{subject to: }& (t_1,t_2,t_3)\in D_{\text{Bell}}\\
&&t^2_1+t^2_2+t^2_3=4p_2-1\in[0,\tfrac13].
\end{eqnarray*}
Due to the fact that the whole sphere
$S:t^2_1+t^2_2+t^2_3=4p_2-1\in[0,\tfrac13]$ is contained in
$D_{\text{Bellsep}}$. By using Lagrange's multiplier method, we
define Lagrange function
\begin{eqnarray*}
L(t_1,t_2,t_3,\lambda):=
t_1t_2t_3+\lambda(t^2_1+t^2_2+t^2_3-4p_2+1),
\end{eqnarray*}
defined over the 1st orthant by symmetry. Then
\begin{eqnarray*}
\begin{cases}
\frac{\partial L}{\partial t_1} &= t_2t_3+2\lambda t_1=0 \\
\frac{\partial L}{\partial t_2} &= t_1t_3+2\lambda t_2=0\\
\frac{\partial L}{\partial t_3} &= t_1t_2+2\lambda t_3=0\\
\frac{\partial L}{\partial \lambda}&=t^2_1+t^2_2+t^2_3-4p_2+1=0
\end{cases}
\end{eqnarray*}
leads to the solution $t_1=t_2=t_3=\sqrt{\frac{4p_2-1}3}$, the
extremal (maximal) point at which $p_3$ attains its maximal value
\begin{eqnarray*}
p_3=\frac{1+6\Pa{\sqrt{\frac{4p_2-1}3}}^3+3(4p_2-1)}{16} =
\Phi^+_4(p_2).
\end{eqnarray*}
In summary, this is the desired result: $\Phi^-_4(p_2)\leqslant
p_3\leqslant\Phi^{\B}_4(p_2)$, where
\begin{eqnarray*}
\Phi^{\B}_4(p_2) =
\begin{cases}
\Phi^+_4(p_2),&\text{if }p_2\in[\frac14,\frac13],\\
\frac{(18p_2-1)+2\sqrt{2}(3p_2-1)^{\frac32}}{36},&\text{if }p_2\in[\frac13,\frac12],\\
\frac14,&\text{if }p_2\in[\frac12,1].\\
\end{cases}
\end{eqnarray*}

\end{document}